# WORK BREAKDOWN STRUCTURE: A TOOL FOR SOFTWARE PROJECT SCOPE VERIFICATION

Robert T. Hans

Software Engineering Department, Tshwane University of Technology, Pretoria, South Africa
hansr@tut.ac.za

*ABSTRACT*

*Software project scope verification is a very important process in project scope management and it needs to be performed properly and thoroughly so as to avoid project rework and scope creep. Moreover, software scope verification is crucial in the process of delivering exactly what the customer requested and minimizing project scope changes. Well defined software scope eases the process of scope verification and contributes to project success. Furthermore, a deliverable-oriented WBS provides a road map to a well defined software scope of work. It is on the basis of this that this paper extends the use of deliverable-oriented WBS to that of scope verification process. This paper argues that a deliverable-oriented WBS is a tool for software scope verification*

## 1. INTRODUCTION

[1] states:
> '*effective scope management is one of the key factors determining project success*'.

This is so because a software project manager's success in managing a project is primarily based on how well he/she manages the project's triple constraint (project scope, cost, and time). User's needs must be well interpreted in the process of scope definition. The process of ensuring that a project scope is well defined and covers all user requirements aspects has proven to be a challenge for software project managers. Both [1] and [2] agree with this assertion and state that one of the critical and difficult aspects of project scope management has been project scope definition. Furthermore, [3] also states that understanding of the project scope (the work that needs to be done) has continuously been the Achilles' heel of project management. It is important that a project team should have a well defined project scope right from the start of a project in order to:

- minimize project scope changes and thereby influencing factors which cause scope changes and aiding in scope control,
- minimize project scope creep,
- ensure that products produced by the project meet user requirements and
- avoid unnecessary project rework
- avoid cost and time overrun.

The key question is how best can software project managers produce a project scope that meets all the abovementioned requirements? This is possible through the scope verification process. [2] defines scope verification as a process that:
> '*involves formal acceptance of the completed project scope by the stakeholders*'

In order to do a thorough job in project scope verification, project team members and users should work closely together throughout the project life cycle. Moreover, [2] states:

> *'The **main tool** for performing scope verification is **inspection**. The customer, sponsor, or user inspects the work after it is delivered'.*

The field of software engineering has numerous research studies and books that have been written on the subject of scope verification. However, the researcher is not aware of any research study on using work breakdown structure (WBS) to verify software project scope. Therefore, this research paper seeks to close this theoretical void by suggesting the use of WBS as a tool for verifying software project scope. This aim is also strengthened by [1] and [4] who state that a WBS provides a road map to a well-defined scope of work and is an important tool for scope quantification. It is on the basis of these assertions that this paper extends the use of WBS to that of scope verification process.

## 2. LITERATURE REVIEW
### 2.1 Introduction

Software project success is generally measured using the triple constraint ('iron/golden' triangle), namely, cost goals, time goals and project performance in terms of project scope completion [5]. Therefore, an organization would want to have a software project scope finished within these constrains. Software companies are in constant search for tools and techniques that would assist project managers manage the triple constraint. Project scope management is one of the knowledge areas in which a project manager should have adequate skills competencies. Moreover, project scope management ensures that project stakeholders have the same understanding of the type of products the project will produce [2]. This is achieved through the use of scope definition and scope verification processes which are part of project scope management. The next section discusses the scope verification process.

### 2.2 Project Scope Verification

In broad terms, verification can be regarded as a 'process of determining whether a software development phase has been correctly carried out' [6]. On the other hand [3] defines verification as a process used to formalize project scope acceptance. Scope verification is a process that is carried out under specification analysis phase. In this phase client's requirements are analyzed and interpreted in order to produce a specification document which will contain project scope of work. Project scope verification is an important process in ensuring that the project team delivers exactly what the customer requested [5] and also in ensuring that project scope changes are minimal [2]. It is the process that formalizes the acceptance of the project scope. On the other hand, [3] regards verification as more than just a process which ensures deliverables conform to user requirements, but also as a quality assurance process too – ensuring that project work is complete and correct.

Various tools and techniques have been suggested for project scope verification process [2], [6] and these include:
- Inspection – the customer or user inspects the work after it is delivered.
- Prototyping – working replica of the planned system.
- Use case modeling – a tool used to model business events in order to gain better understanding of user requirements.
- Joint application design – a technique used to promote greater involvement of key project stakeholders in system development.
- Walkthrough – A document is carefully checked by a team of software professionals.

Any of these tools and techniques may be applied in the process of project scope verification. In order for project scope verification to proceed smoothly, scope definition has to be done thoroughly. A well defined project scope is a pre-cursor for successful project scope control, project success and customer satisfaction. The next section discusses work breakdown structure which plays an important role in defining a project scope.

## 2.3 Work Breakdown Structure

According to [2] a work breakdown structure is:
> 'a deliverable-oriented grouping of the work involved in a project that defines the total scope of the project'.

It is a hierarchical structure which is normally represented in a graphical form [1], [3] or in a tabular form. The graphical form is appropriate for communicating work activities to both top management and/or customers, while the tabular form is useful for cost and schedule development [3]. In essence, a WBS is a decomposition of a project scope into smaller manageable parts. The decomposition can proceed until the desired level of detail, lowest-level discrete deliverables, is reached [7], [8]. A WBS defines the total project scope [2], [5], [9].

A work breakdown structure is important to the success delivery of any software project [7] because it defines in detail the work necessary to accomplish a project's objectives. It also shows interim deliverables required to produce major project deliverables stated in project scope definition [3]. The successful delivery of the software project is in terms of cost and schedule [10] performance – a WBS provides a foundation for better cost and schedule estimation. [11] agrees with this assertion and states that a WBS is widely regarded as powerful tool for better performance control. Furthermore, a WBS is backbone of the project and without it a project manager would be attempting to manage a shapeless project [1]. A typical WBS reflects user and/or system requirements [7] as well as providing a basis for identifying resources and tasks for developing a project cost estimate. A well developed WBS also serves as a communication tool among all stakeholders [12], [3]. [4] agrees with this assertion and states that a good WBS reduces the possibility of omitting key project elements. Software project managers use various tools and techniques in developing a WBS. The next section briefly discusses different approaches used in WBS development.

## 2.4 Approaches in Developing WBS

The primary technique in creating a WBS is the decomposition process [7], [2]. A WBS can be presented in two forms, namely, hierarchical and tabular form [7]. According to [2] there are various approaches that can be used in developing a WBS and these include:
- Using guidelines – organizational guidelines that are to be followed in developing work breakdown structures in the organization.
- An analogy approach – the use of a similar project's WBS in developing a WBS of the new project.
- The Top-down approach – one starts with largest items of the project and break them down into their subordinate items.
- The Bottom-up approach – one identifies specific related project tasks and organize them into summary activities (aggregate tasks).
- Mind mapping approach – one starts with a core idea and then link related ideas to the core idea.

There is no single best approach in developing a WBS and therefore it is possible to use a combination of approaches in a WBS development [2].

According to [13] in recent years experts have considered another approach to project planning and have suggested the use of product breakdown structure (PBS) or a deliverable-oriented WBS [4] in creating a WBS. The advantage of this approach is that the project's focus is on what is to be achieved rather than how, in other words the focus is on the ends instead of the means [13]. The creation of a PBS uses the similar principles as in the creation of a WBS – progressively decomposing the project products into smaller products until a sensible, unitary product level is reached. The aim is to create a WBS which highlights a logical organization of products, parts and modules [4]. The next section discusses the role that a deliverable-oriented WBS can play in scope verification.

## 3. THE IMPORTANCE OF DELIVERABLE-ORIENTED WBS IN SCOPE VERIFICATION

[14] emphasizes the importance of having the end results in mind right from the beginning of a project. The deliverable-oriented approach in creating a WBS enforces this – the end result is project deliverables. According to [4] a deliverable-oriented WBS provides a road map to the definition of scope of work. It also provides a 'clear picture of what needs to be accomplished' [12]. Furthermore, deliverables serve as a basis for creating a WBS [4]. Moreover, a deliverable-oriented WBS facilitates the sharing of information and serves as a tool for communication and project scope control [8]. The process of scope verification involves

establishing as to whether the developed scope does conform to the user requirements [6] – verifying scope against user requirements, as indicated by Figure 1 below. That is, verifying that the scope addresses all user requirements. Therefore, scope verification encompasses, in a way, requirements analysis.

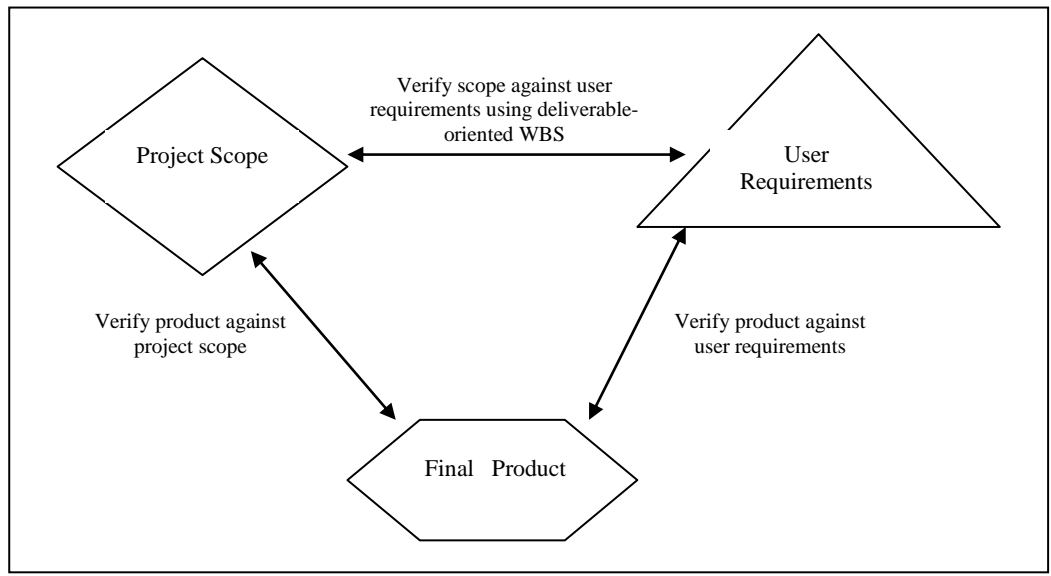

Figure 1. Scope, User requirements and Final product verification process

In the process of clarifying the project scope and establishing its completeness with the users, one would be able to find out if the product(s) that will be produced through the completion of project scope would meet user requirements [15]. A project scope which is contained in the project specification document [6] is a product of user requirements analysis process. Moreover, a deliverable-oriented WBS, which is the total project scope [3], focuses on what is to be produced in terms of project products. Furthermore, a good WBS entails all the work of the project and all project deliverables are explicit in it [3]. Therefore, such a structuring of work (scope) makes it easier for project managers to verify the scope against user requirements, because he/she knows what deliverables are expected in order to meet user requirements. The primary purpose of scope verification as it was pointed out above, is to establish the completeness of project scope in terms of user requirements. This would entail establishing as to whether all scope deliverables as outlined by the WBS are in fact complete in terms of user specification. This cross checking or reference using a deliverable-oriented WBS would also help detect misinterpretation of user requirements as well as omissions that might have occurred in the process of specification document development.

Scope verification against user requirements is important and should be done thoroughly in order to minimize the impact that would result from final product verification against project scope as well as final product verification against user requirements. Should the scope verification against user requirements be not done well, the other two verifications could result in project rework which would be costly as the final product has already been produced at that stage.

## 4. CONCLUSION

In the light of avoiding software project scope rework and promotion of better project scope control, project scope verification process is critical. Moreover, software scope verification is an important process in ensuring that the project team delivers exactly what the customer requested and minimizes project scope changes. A deliverable-oriented WBS provides project managers with a scope verification tool to ease the challenge of scope verification [2]. This paper has argued this assertion in the preceding sections.

## 5. LIMITATIONS

This paper has only presented a conceptual argument in as far as using a deliverable-oriented WBS for software project scope verification process, and therefore empirical evidence is needed to support the claim.

## 6. FUTURE RESEARCH

The empirical usefulness and benefits of using a deliverable-oriented WBS as a software project scope verification tool need to tested and validated.

## 7. ABOUT THE AUTHOR

Full name:            Mr Robert Hans

Affiliation:           Member of IEEE

E-mail address: hansr@tut.ac.za

Full international contact details:

Telephone: +27 (0) 12 382 9721

Fax: +27 (0) 12 382 9203

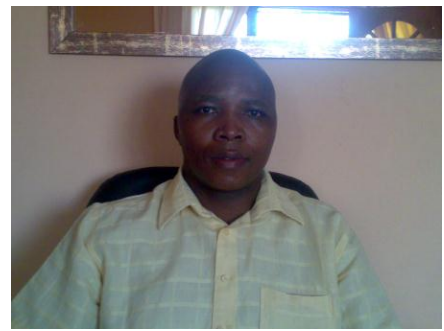

Brief professional biography:

Robert Hans is a lecturer at Tshwane University of Technology (TUT) in the department of Software Engineering, where he teaches project management as well as systems analysis and design subjects. He has worked for many years in the ICT industry before joining TUT in 2008. Robert has a masters degree in business leadership (MBL) from University of South Africa Graduate School of Business Leadership (Unisa-

SBL). He received a distinction for his research paper in his MBL degree. He is currently studying for a doctoral degree in computer science with the University of South Africa. So far Robert has published four peer-reviewed conference papers and one peer-reviewed journal paper in project management.